\documentclass[10pt]{article} 

\usepackage{hyperref}

\usepackage{geometry} 
\usepackage{graphicx} 
\usepackage{amsmath}
\usepackage{epstopdf}

\title{A New Tracking Algorithm for Multiple Colloidal \\ Particles Close to Contact}
\author{Harun Y\"{u}cel\footnote{Corresponding author: harun.yucel@oposta.omu.edu.tr}\phantom{1} and Nazmi Turan Okumu\c{s}o\u{g}lu \\ \\
Department of Physics, Faculty of Arts and Sciences, \\ Ondokuz May{\i}s University, 55139, Samsun, Turkey
}
\date{}
\begin{document}
\maketitle

\noindent {\bf This document is a preprint version: for the (revised) accepted version, please refer to the publisher's website via the following \\ J. Phys.: Condens. Matter, 2017, DOI : \href{https://doi.org/10.1088/1361-648X/aa908e}{10.1088/1361-648X/aa908e}}

\begin{abstract}
In this paper, we propose a new algorithm based on radial symmetry center method to track colloidal particles close to contact, where the optical images of the particles start to overlap in digital video microscopy. This overlapping effect is important to observe the pair interaction potential in colloidal studies and it appears as additional interaction in the measurement of the interaction with conventional tracking analysis. The proposed algorithm in this work is simple, fast and applicable for not only two particles but also three and more particles without any modification. The algorithm uses gradient vectors of the particle intensity distribution, which allows us to use a part of the symmetric intensity distribution in the calculation of the actual particle position. In this study, simulations are performed to see the performance of the proposed algorithm for two and three particles, where the simulation images are generated by using fitted curve to experimental particle image for different sized particles. As a result, the algorithm yields the maximum error smaller than $2nm$ for $5.53\mu m$ silica particles in contact condition.
\end{abstract}


\section{Introduction}
Particle tracking is a technique commonly used to obtain information about micro and nano scale world. Especially, the measurement of the forces to understand micro world dynamics depends on the positions of the observed objects. In digital video microscopy including an optical tweezer, tracking a particle suspended in a liquid is very important because an external force applied to the trapped particle is observed by means of measuring only the particle displacement from its equilibrium position \cite{florin1997, neuman2004}. Moreover, holographic optical trapping (see \cite{grier2006} for example) allows us to trap and to observe multiple colloidal particles, where both pair interaction potential between two particles (see \cite{crocker1994} for example) and many body interaction among many particles (see \cite{brunner2004, Jan2009, Sathya2016} for example) can be studied especially at interparticle distances. Therefore, it is needed high precision methods that can track particles close to contact.

In literature, there are many position detection algorithms. The centroid method is well known algorithm and the developed version of this algorithm is published in \cite{Jer2015}. The other commonly used algorithm is Gaussian fitting method \cite{Croc1996} which gives more accurate results than the centroid methods. Recently, the radial symmetry center method is developed in \cite{Rag2012,Lin2013,Ma2012,Ma2015} to determine the position of a particle under digital video microscope. This method is faster than Gaussian fitting method and it can detect the position of the particle more accurately.

These methods assume that the maximum intensity point in the particle image corresponds to the particle position. Thus, the algorithms have limitation when two particles are close each other in observing pair potential between particles \cite{Bech2005, Mar2007, ramirez2006, Gyg2008}. In this condition, the diffraction patterns of the particles in the image start to overlap, which cause the maximum intensity points of the particles shifted. Therefore, tracking particles close to contact by using conventional algorithms gives unreal interaction potentials \cite{Bau2006}. The overlapping effect in the image of the particles is a problem for not only two particles but also three and more particles. 

In the literature, one of the methods solving the overlapping problem is correction method \cite{Bech2005}, where a correction function is obtained by using simulated images. In the simulation, an arbitrary  function fitted to experimental image is used to generate one particle intensity distribution and two particles images are obtained by linear combination of the fit functions. Therefore, the positions of the particles obtained by using conventional methods are corrected. This technique takes more time and it is not easy to apply for three and more particles because it is difficult to obtain the correction function for three particles.

The other study about solving the overlapping problem is presented by Zhang et. al. \cite{Zhang2015}, where C-shaped template based tracking algorithm is used as a base algorithm. C-shaped template provides tracing only a fraction of the particle image. Their algorithm suggests using undistorted part of the particle intensity distribution in the image. However, the algorithm needs the orientation information between the particles and it is difficult to adapt it for more particles. Moreover, their algorithm includes to track an isolated reference particles to reduce systematic errors \cite{Zhang2015, Har2017}.

In the literature, there is a few new algorithms \cite{Ueb2012, Raud2015, Wel2017} solving the overlapping problem for different applications. Those algorithms use model fitting where the experimental system should be characterized in terms of intensity distributions. Thus, the algorithms are complex and their use is troublesome. 

In this paper, we propose a new algorithm based on radial symmetry center method \cite{Rag2012,Ma2012} to solve the overlapping problem. Radial symmetry center method uses gradient vectors of the intensity distribution of the image. The gradient vectors intersect in the center of the particle intensity distribution. The method can easily and quickly find the intersection point by means of analytical solution. Here, we focus on to use a fraction of the particle intensity distribution for radially symmetric distribution. The gradient vectors allow us to apply it as shown in \cite{Har2016} and it provides an advantage to solve the overlapping problem in the particles images. If the particle intensity distribution covers only a several ten pixels area, using a part of the distribution can be useless. But the distribution within several hundred pixels area can provide using a part of the distribution to obtain the particle center coordinates. Our algorithm uses the undistorted part of the particle image to determine its position and the proposed algorithm is applicable for any number of particles tracking without any modification.

In this study, we performed simulations for two and three particles to demonstrate the performance of the proposed algorithm. In simulations, we used a fitted curve to experimental particle image to generate particle images. As a result, we can conclude that our algorithm gives as good results as available methods and it is useful for many body interaction experiments at interparticle distances.

\section{Materials and Methods}
\begin{figure}[t]
\begin{center}
\includegraphics[width=5cm]{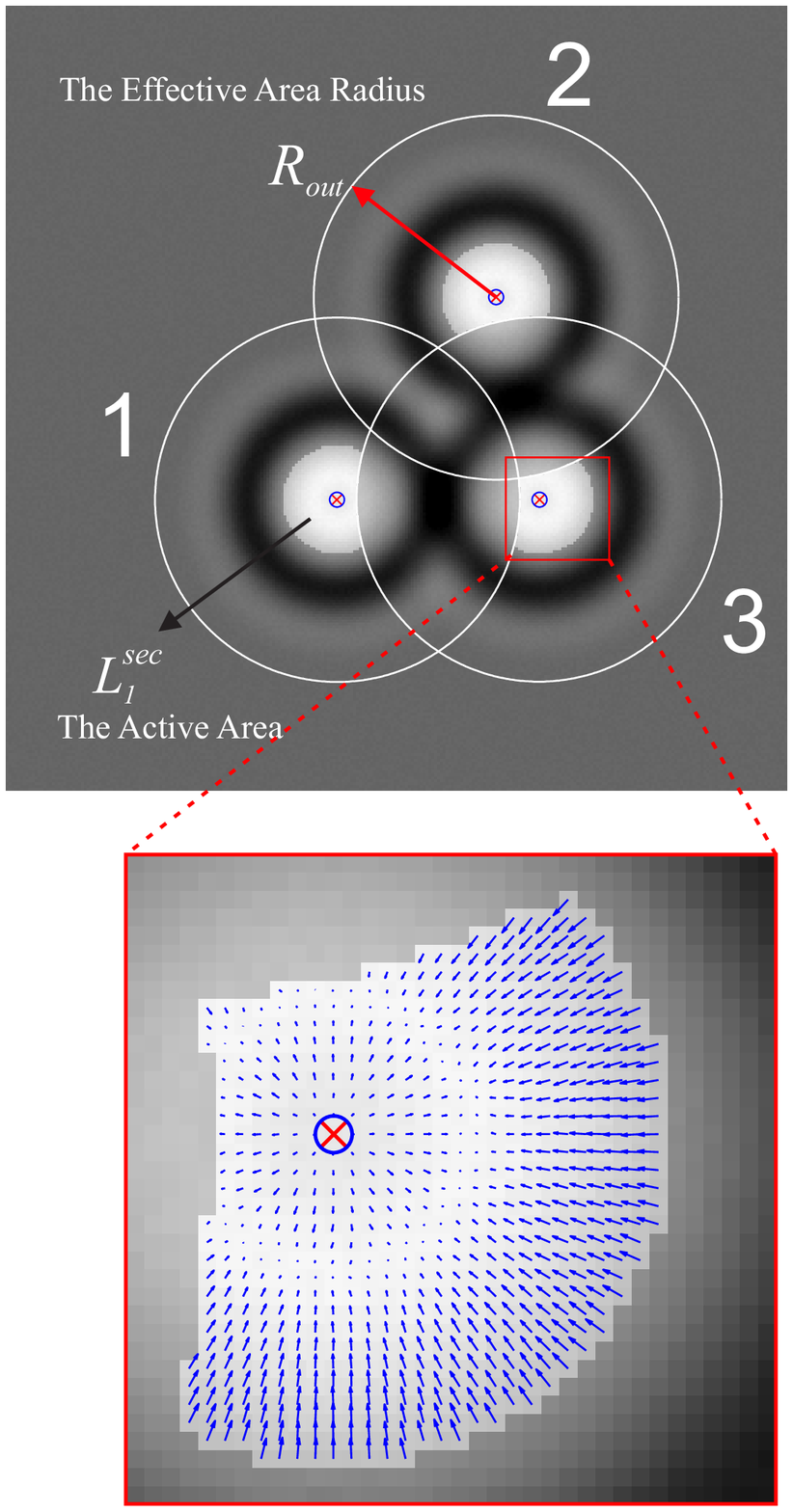}
\end{center}
\caption{\label{fig1} A presentation of the algorithm. The gray regions correspond to the seconder active areas (undistorted area) where the gradient vectors are used in the calculation of the centers. The white rings show the effective area borders for each particle. In the bottom, the gradient vectors are shown within the active area for 3th particle, where the gradient vectors intersect in the center points.}
\end{figure}

In this study, we consider radial symmetry center method proposed in \cite{Rag2012,Ma2012} to find the position of the particle in the camera image $I$ which can be seen in Fig \ref{fig1}. Radial symmetry centers method uses the gradient vectors of the image intensity distribution and the intersection point of the vectors presents the intensity center of a symmetric intensity distribution. Thus, the determination of the gradient vectors is important and their directions are affected by the acquisition noise in the image. Therefore, we focus on the bright region of the particle intensity distribution where the signal is dominant and we use a threshold value $th$ to determine the bright region which is called as active area of the particle through this work. Addition, this approach provides to determine the active areas in case of multiple particles condition in the image. In the calculation of the center, we use the intensity information within the active area $L$ 
\[
L(i,j) = \begin{cases}
    1  & I(i,j) \geq th\\
    0  & I(i,j) < th\\
  \end{cases}
\]
Here $i=1,2,3, \dots, N$ and $j=1,2,3, \dots, M$ are row and column indexes of the images. For multiple particles, we use labeling algorithm from \cite{Hera1992} to obtain the active areas for each particle. The labeling algorithm uses the binary image and gives numbers for each particle region like $k=1,2,3, \dots$. Therefore, we can extract binary images for each particle and obtain the active area arrays $L_k$ which have the same size with image $I_{NxM}$ for each particle. 

To obtain the intensity gradient vectors of the image, we use below gradient operators as suggested by Ma et.al. \cite{Ma2012}. 
\[
g_x=
\begin{bmatrix}
    1 & 1 & 0 & -1 & -1 \\
    1 & 1 & 0 & -1 & -1 \\
    1 & 1 & 0 & -1 & -1 
\end{bmatrix}
* I
 \quad , \quad
g_y=\begin{bmatrix}
    1 & 1 & 1  \\
    1 & 1 & 1  \\
    0 & 0 & 0  \\
    -1 & -1 & -1 \\
    -1 & -1 & -1 
\end{bmatrix}
* I
\]

Here, the operator $*$ means convolution. These operators produce gradient vectors for the midpoint of the image pixels. To find the intersection point of the gradient vectors we must calculate the 
slopes of the gradient vectors which are presented by
\[m(i,j)=g_y(i,j)/g_x(i,j)\]
We denote that $(x_k, y_k)$ is the brightness center of the image $I$ for $k$th particle. If $(x_k, y_k)$ is a center point, the lines with slope $m(i,j)$ should intersect on the center point as can be seen in Fig. \ref{fig1}. This means that when we calculate the perpendicular distance $d$ between the center point and the direction of a gradient vector, the distance should be zero. Thus, total distance between the lines and the center point $(x_k, y_k)$ should be minimum because of noise fluctuations in the image \cite{Rag2012, Ma2012}. The distance $d(i,j)$ from $(x_k, y_k)$ center point to the lines is given by following form
\[
d(i,j)^2 = \frac{\left [ (i-y_k) + m(i,j)(x_k-j)  \right ] ^2}{m(i,j)^2+1}
\]
and total distance can be calculated as
\begin{equation} \label{eq2}
D^2_k = \sum_{i,j} w_k(i,j) d(i,j)^2
\end{equation}
Here, considering the square of the total distance is useful to able to solve analytically Eq. \ref{eq3}. A weight function $w_k(i,j)$ can be considered to eliminate useless areas like background, diffraction pattern and gradient vector that is affected by noise in the image. We use the magnitude of the gradient vectors to reduce the effect of the noise, because when the changing of the intensities between pixels are low, the influence of the noise is dominant on the direction of gradient vector. In weight function, the active area array eliminates the background and the diffraction pattern. 
\begin{equation} \label{eq1}
w_k(i,j)=L_k(i,j) \sqrt{ g_x(i,j) ^2  + g_y(i,j) ^2 }
\end{equation}
At position $(x_k, y_k)$, the total distance $D^2_k$ takes minimum value.
\begin{equation} \label{eq3}
 \begin{cases}
    {\partial D^2_k}/{\partial x_k} = 0\\
    {\partial D^2_k}/{\partial y_k} = 0\\
  \end{cases}
\end{equation}
Therefore, we can calculate the center coordinates $(x_k, y_k)$
easily by solving the linear equations in Eq.\ref{eq3} for each particle \cite{Rag2012,Ma2012}. The solution is given in Appendix \ref{app}.

We call this center coordinates as first estimated centers which have systematic errors when particles are close to contact because of the overlapping effect. To reach the actual centers, we focus on the undistorted part of the intensity distribution of the particle. Theoretically, the gradient vectors in the undistorted part of the intensity distribution still intersect on the actual center if the individual particle intensity distribution is radially symmetric. Thus, we must predict the undistorted regions of the particle intensity distributions on the image in close contact condition.

For reaching this aim, we follow an easy way. First, we describe the effective area of the individual particle, where the effective area is a circle with radius $R_{out}$ which covers all particle intensity distribution including its diffraction pattern as can be seen in Fig.\ref{fig1}. Here, we can generate a new binary array to describe the effective area $L_k^{eff}$ by means of the parameters $R_{out}$ and the first estimated center coordinates. 
\begin{equation}
L_k^{eff}(i,j) = \begin{cases}
    1  & \sqrt{(j-x_k)^2 + (i-y_k)^2} \leq R_{out}\\
    0  & \sqrt{(j-x_k)^2 + (i-y_k)^2} > R_{out}\\
  \end{cases}
\end{equation}
When the active area $L_k$ of the one particle and the effective area $L_h^{eff}$ of the other particle start to overlap, we can approximately predict the undistorted part of the particle intensity distribution by calculating, second times, the second active areas $L_k^{sec}$ as gray pointed areas in Fig.\ref{fig1}.
\begin{equation} \label{eq4}
L_k^{sec}(i,j)= \begin{cases}
    1  & L_k(i,j) + \sum_{h\neq k} 2L_h^{eff}(i,j) = 1\\
    0  & L_k(i,j) + \sum_{h\neq k} 2L_h^{eff}(i,j) \neq 1\\
  \end{cases}
\end{equation}
Then we recalculate Eq.\ref{eq1}, Eq.\ref{eq2} and Eq.\ref{eq3} by using $L_k^{sec}$ instead of $L_k$ as it is in Fig.\ref{fig1} and we obtain the actual center coordinates. We present the algorithm in the code file (framedetec\_ef.m) in the supplementary file.

In this approach, the second active area $L_k^{sec}$ may still have a little distorted area causing a little systematic error in the calculation of the center coordinates. Because the determination of the effective area $L_k^{eff}$ is based on the first estimated centers which have systematic errors when particles are close to contact. However, this way allows the elimination of a major part of the distorted areas, which improves accuracy of calculated center coordinates.  At this point, the character of the diffraction pattern of the particle becomes important in terms of errors. A short range and a soft fluctuation of the diffraction pattern around the particle are suitable for the success of the algorithm. A soft fluctuation according to background intensity provides the errors to be smaller than a few pixels and the effective area can be determined at high accuracy. Thus, if the actual centers are close to the first estimated centers (it shows intensity center) by a few pixels as observed in experiments performed in \cite{Bech2005, ramirez2006} with standard equipment, we can reach the accurate results. A sort range of the pattern provides to use an acceptable active area on the other particle. If there is no undistorted area on the particle image, the algorithm cannot yield a result.

In two-particle condition, the mass center can be used a reference point to determine distorted areas by using particle diameter $D$ and the range parameter $R_{out}$. On the axis of the first estimated centers, we can determine the distorted area by drawing the intersection region of two circles with radius $R_{out}$, where their centers are placed $D/2$ away from the mass center in opposite directions. The intersection region of the circles shows the distorted area when particles in the contact condition. This is a complete solution for two-particle condition. But for three and many particles, it is not a full solution. Therefore, we consider the former approach.

\begin{figure}[!t]
\begin{center}
\includegraphics[width=8cm]{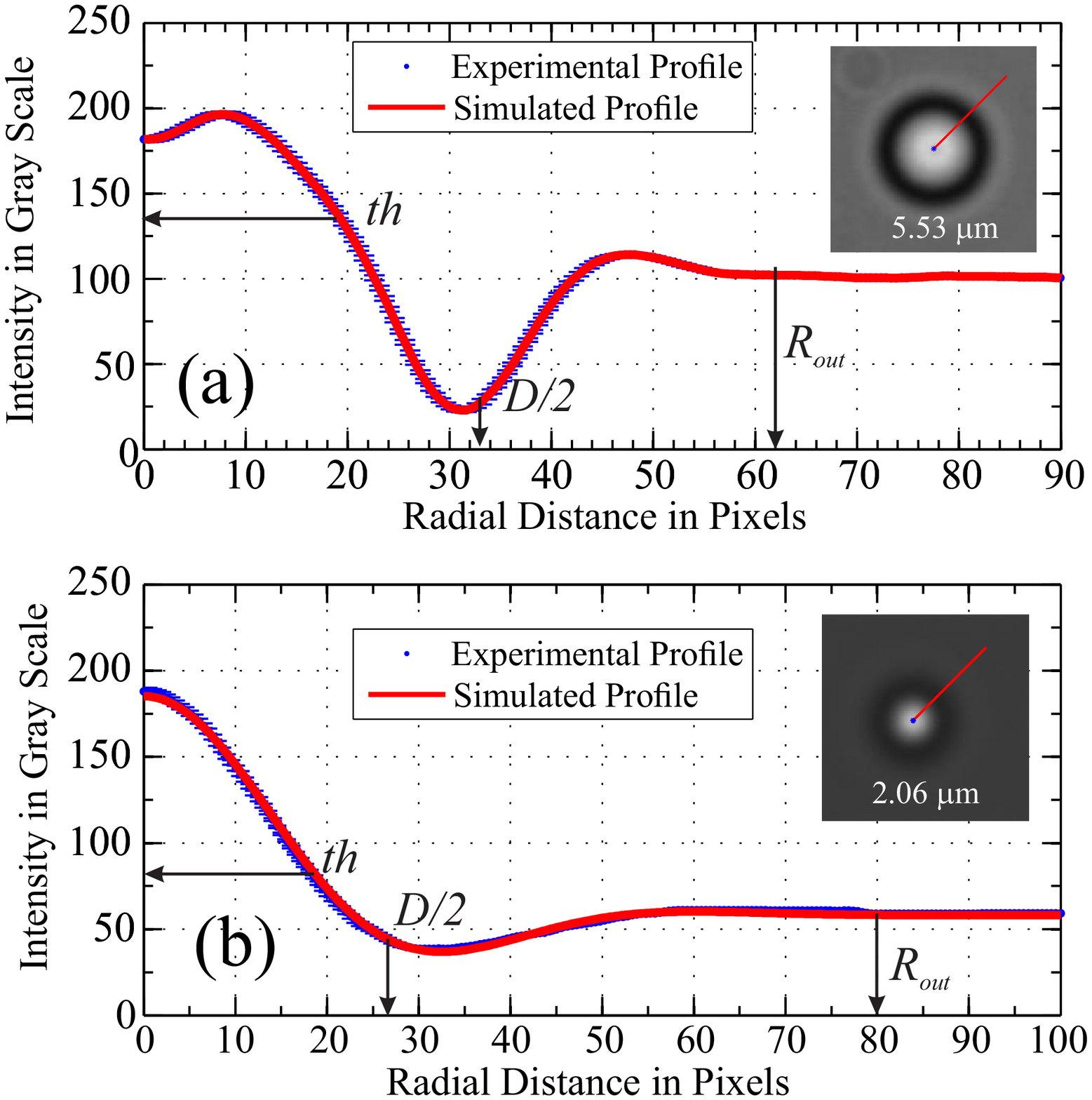}
\end{center}
\caption{\label{fig2} The intensity distribution profiles of particle images. (a) is for $5.53 \mu m$ and (b) is for $2.06 \mu m$ diameter particles. In the graphics, it can be seen both the experimental and the simulated profiles and $R_{out}$, $D/2$ and $th$ values are marked for each particle. The insets of the graphics show the experimental particle images taken by digital camera.}
\end{figure}

In the application of the algorithm, the parameters $R_{out}$ and $th$ can be easily determined by analyzing the intensity profile of the individual particle as shown in Fig.\ref{fig2}. $R_{out}$ can be the transition point where the intensity of the particle is ended and the background intensity is started in the intensity profile of the particle. Choosing $R_{out}$ little higher than the transition point provides the consideration of the little fluctuations of the particle intensity distribution due to the small displacement of the particle along the axis of the optical system. Similarly, $th$ threshold value can also be determined by choosing a suitable value to extract the central intensity area, the active area, of the particle image in contact condition.

In this paper, we tested the proposed algorithm by using simulated images. For single particle image simulation, the intensity profile can be calculated numerically based on the general Mie theory \cite{Ovr2000}. However, this approach needs many experimental parameters. Thus, we used the particle intensity profile obtained from the experimental image. First, we considered the experimental image which is presented its image and intensity profile in Fig. \ref{fig2} (a). In the experimental system, the images of the silica particles with diameter $D=5.53\mu m$ were taken by a digital camera under white light (incoherent) illumination with $63x$ ($NA = 0.9$) Zeiss microscope objective. In the system, we found that one pixel corresponds to $0.0847 \mu m$. After the extracting of the intensity profile of the particle $I_{exp}$, we generated the simulated image by using interpolation of the experimental profile as formulated $I_{1p} = I_{exp}+I_0$, where $I_0$ is the background intensity.

For a better understanding of the performance of the algorithm, we examined it by using $2.06 \mu m$ sized silica particle images which are inconvenient for the algorithm. The images were taken under white light (incoherent) illumination with $100x$ ($NA = 1.3$) oil immersion Zeiss microscope objective. The image and its intensity profile are given in  Fig. \ref{fig2} (b) and its simulated images were generated by using the fitted curve (a Gaussian-Bessel function) to experimental intensity profile instead of interpolation. In the system, the conversation parameter was found as $0.037 \mu m$.

In the generation of many particle images, we used a linear combination of each particle profile as explained in \cite{Bech2005,ramirez2006,Gyg2008}. The formulation can be given as   $I = I_{exp}^{1p}+I_{exp}^{2p}+I_0$. This approach may not be always appropriative especially for the middle region between particles. But the proposed algorithm here doesn't use the information in the middle region. For noise generation in the images we used randomly fluctuations which its amplitude is $5$ unit in gray scale.

\begin{figure}[!b]
\begin{center}
\includegraphics[width=6cm]{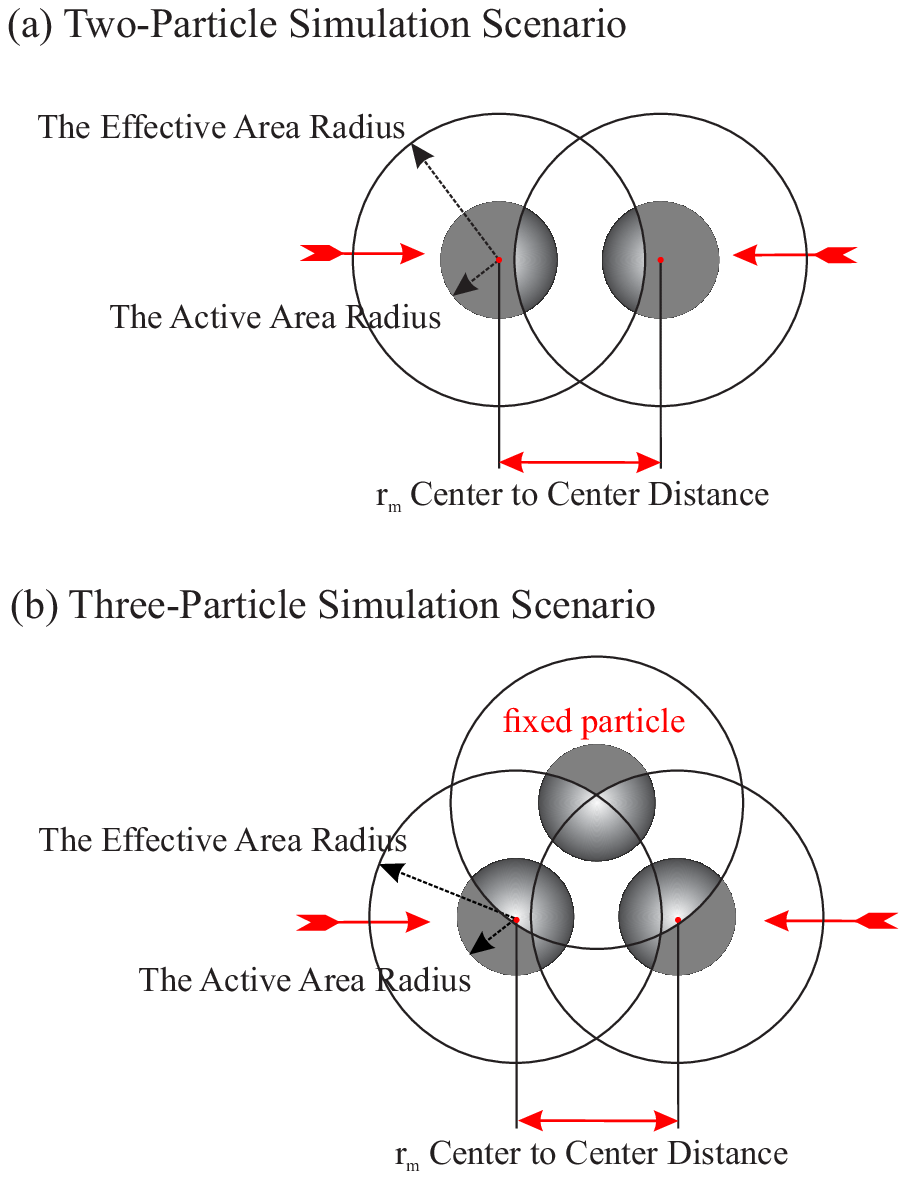}
\end{center}
\caption{\label{fig3} (a) shows the scenario of the simulation for two particles and (b) presents the simulation scenario for three particles. The red arrows show the motion of the particles and center to center distance is observed. The gray areas correspond to the active areas which are used to calculate the accurate position.}
\end{figure}

\section{Results and Discussions}
In simulations, we consider two simulation scenarios which are presented in Fig. \ref{fig3}. One of them, (a), is a two-particle system where the relative distance $r_m$ is observed between the particles. The other scenario, (b), considers a three-particle system where we observe the relative distance $r_m$ between the two particles that the intensity distribution of third particle creates additional image distortion on them. The proposed algorithm describes automatically the particle number and determines the active areas (gray regions in Fig. \ref{fig3}) to calculate the accurate particle centers.

In two particles simulation, we generated the particle images with $r_t$ separation distance which is changed from non-overlap condition of the diffraction patterns to contact condition of the particles step by step. We used the experimental system parameters to determine the minimum (contact) distance. For each step, we regenerated the image two hundred times to evaluate statistical errors. Then, we determined the relative distance $r_m$ from the generated images and calculated the absolute errors by using different position detection algorithm. We performed two-particle scenario for $5.53 \mu m$ and $2.06 \mu m$ particle images.

\begin{figure}[!b]
\begin{center}
\includegraphics[width=8cm]{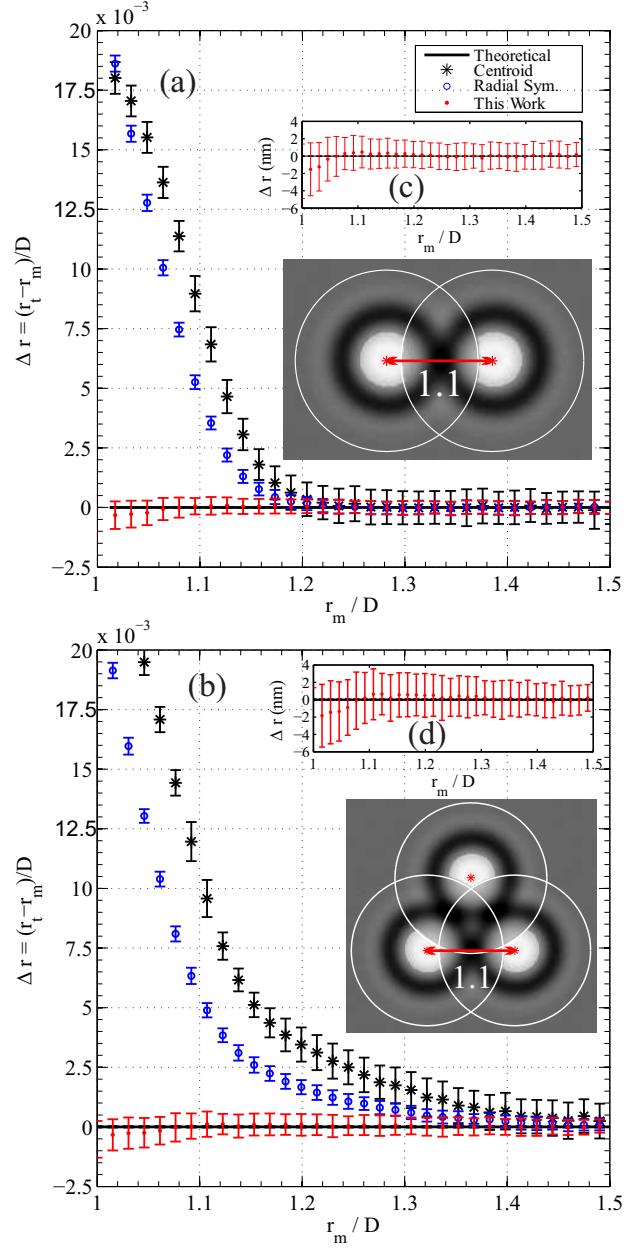}
\end{center}
\caption{\label{fig4} For $5.53\mu m$ sized particle, the graphics (a) and (b) show the results of the simulations for two particles and three particles, respectively. The graphics c) and d) present the results of this work in nanometer unit. The insets of the graphics show the images of the particles where the active areas of the particles are presented for $1.1$ relative distance during simulations.}
\end{figure}

Three particles simulation is very similar the two particles simulation. We launched the simulation in condition that three particles are in contact with equilateral triangle position, as can been seen from Fig \ref{fig4} (b). We kept constant the position of the middle particle and have increased the relative distance between the other two particles step by step. As with two particles simulation,     we regenerated the image two hundred times to evaluate statistical errors for each step and determined the relative distance $r_m$ from the generated images. We performed three-particle scenario for only $5.53 \mu m$ particle images.

For $5.53 \mu m$ particle, the simulation results are given in Fig. \ref{fig4}. The graphics (a) and (b) are for two and three particles simulations, respectively, and the particle images taken during the simulations are shown in the graphics. In Fig. \ref{fig4}, the graphics (c) and (d) present the results of the proposed algorithm in nanometer unit. As can be seen from Fig. \ref{fig4}, when the particles are close to contact, even contact condition, the proposed algorithm yields more accurate relative distance. In average, the proposed algorithm yields $2nm$ maximum errors for both simulations in contact condition. When the particles are closer each other, the absolute errors of the proposed algorithm increase by a very small amount. The reason for this may be that the second active area $L_k^{sec}$ has still a little bit distorted area or errors coming from image generation process. But this increasing is very small comparing to the errors of the other methods. An important point noticed from the images in the graphics is that the active areas of the particles (gray painted regions) have almost half of the full areas at the contact condition.

From the graphics in Fig. \ref{fig4}, we can see that the statistical errors in three particles results are bigger than those of two particles results and this is an expected result. Because in three-particle condition the active area used in the calculation of the actual centers are smaller than those of two-particle condition. Therefore, we can conclude that the smaller active area yields the higher statistical errors.

\begin{figure}[!t]
\begin{center}
\includegraphics[width=8cm]{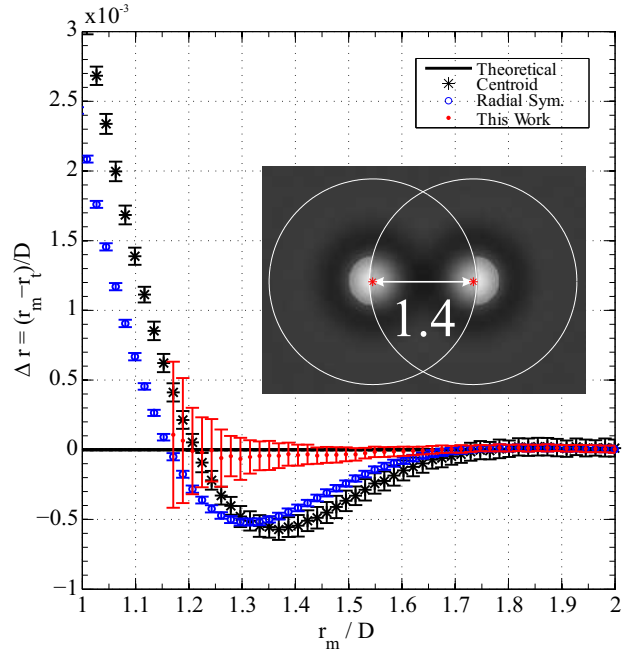}
\end{center}
\caption{\label{fig5} The graphic shows the results of the simulation for $2.06 \mu m$ sized silica particles. The inset of the graphic presents the simulation image where the situation of the active areas can be seen for $1.4$ relative distance.}
\end{figure}

For $2.06 \mu m$ particle, the simulation results are given in Fig. \ref{fig5}. We should emphasize that the intensity distribution of the particle is not convenient for the algorithm due to its long range. When the particles are in the contact condition, there is no undistorted area on the image. Thus, the algorithm cannot yield a result. The other result to be seems from the graphic is the statistical errors of the proposed algorithm. For the relative distance between $1.2$ and $1.4$, the statistical errors increase quickly although the algorithm can estimate more accurate results in average. The reason for this can be the size of the active areas and their positions according to particle centers. The images in the graphic shows the particles positions and their active areas for $1.4$ distance, where the active areas have approximately half of the full areas. In case of the closer distances than that distance, the size of the active areas is smaller and their positions are far from the centers, that is, the active areas don't include the particle centers. This is the main reason that increase the statistical errors. When we consider arbitrary two pixels in an active area which is far from the center, their gradient vectors intersect at the centers. However, the noise on the image causes a little fluctuation on the directions of the vectors, which is the reason why their intersection point has big spatial fluctuation. This effect can be observed in the statistical errors of the all simulations where we have used randomly noise in image generation.

From the results, we can conclude that half of the full areas for the active area yields acceptable center coordinates. This situation can be provided if $R_{out} \leq D$ for contact condition. The intensity center of the one particle should not be affected by the diffraction pattern of the other particle in the contact condition for a two-particle system. Here, $D$ corresponds to the diameter of the particle which determine the contact distance and the size of the active area. $R_{out}$ is a measure of the range of the diffraction pattern. The success of the algorithm needs big size of the active area and a short range of the diffraction pattern. Moreover, a shorter range of the pattern and a bigger size of the active area allow the tracking of many particles in contact condition. For example, if three particles are placed on a line in contact, the middle particle would have the distortion for both side. A short range pattern of the other particles allows us to calculate the center coordinates of the middle particle.  

The used images in the performed simulations have soft fluctuation character. From the graphs for $5.53 \mu m$ , it can be seen that their absolute errors have maximum $1.3$ pixels for the centroid algorithm (or conventional algorithms), where this is calculated by using the conversation parameter and particle's diameter. For $5.53 \mu m$ sized particle, it is easy to simulate images placed particles closer than the contact condition, which is corresponds to unrealistic physical condition. For such simulation, we saw that the proposed algorithm produces the accurate relative distances while the centroid algorithm yields the absolute errors between $2$ and $-4$ pixels corresponding to twice error of a single particle.  
 
Previously, we performed similar simulations with the same experimental parameters for $5.53\mu m$ sized particle in \cite{Har2017} where we aimed to understand the performance of the template based tracking algorithm that are proposed and experimentally verified by Zhang et. al. \cite{Zhang2015}. When we compare the simulation results between two works, we can say that the proposed algorithm in this work is useful in terms of easy applicability and yields more accurate results. When we compare both algorithms, it is seen that unlike the proposed algorithm, their algorithm needs an isolated reference particle to eliminate systematic errors in contact condition.

In this paper, we have generated the simulated images by using the experimental profile of the particle. The image generation process generates only radially symmetric images and the algorithm uses this symmetry. In experimental condition, ones should obtain radially symmetric intensity distribution from the optical setup. Therefore, the shape of the particle should be exactly sphere and bright field illumination with perfectly aligned optical setup is needed. A standard microscope can provide these conditions easily. Additionally, it can be used a noise reduction technique to reduce accusation noise in image processing.

On the other hand, the algorithm can be applied to micro or nano sized fluorescence particle. In this case, the particle intensity distribution can be considered as a 2D radially symmetric Gaussian distribution and the algorithm can produce accurate relative distances within its limitations. By modifying the algorithm, it is possible to apply it to measurement of distances between a particle and a surface. Moreover, the algorithm can be extended for 3D particle tracking by using the proposed algorithm in \cite{Lin2013}.

\section{Conclusions}
In summary, we proposed a new algorithm based on the radial symmetry centers method to determine the positions of the particles close to contact. The algorithm determines the number of the particles and predicts the undistorted region of the particle intensity distribution when the particles are close to contact, automatically. In this study, we have demonstrated by means of simulations that using a part of the symmetric intensity distribution of a trapped particle is useful to track particles close to contact for not only two particles but also three particles. During simulations, we have determined that half of the full active area yields acceptable center coordinates in terms of statistical errors for a two-particle system and we have formulated it as $R_{out} \leq D$ on the image for contact condition. This means the center of the particle should be inside of the active area or very close to the active area. Thus, a short range and a soft fluctuation of the diffraction pattern around the particle are important for the success of the algorithm. On the other hand, the algorithm can be adapted to micro or nano sized fluorescence particles, to measurement distance between a particle and a surface and to 3D particle tracking in contact condition. We expect that the proposed algorithm would provide an important contribution for pair and many body interaction studies at the interparticle distances. One can find Matlab code of the proposed algorithm in the supplementary file. 

\section{Acknowledgments}
We thank to Taha \c Ca\u g\i l (STC Elektronik Ltd.) for the use of their optical equipment.

\appendix
\section{Finding of The Center Coordinates}\label{app}
All formulation is available in ref \cite{Rag2012,Ma2012}. Here we will reproduce it with some modifications for a single particle. The line equation is given by
\[
y_0-y + m(x-x_0)=0
\]
we denote that $i$ is the column index of the image $I$, $j$ is the row index of the image $I$ and $(x_c, y_c)$ is the brightness center of the image $I$, then the line equation can be written as following form
\[
i-y_c + m(i,j)(x_c-j)=0
\]
$d(i,j)^2$ perpendicular distance between $(x_c, y_c)$ center point and line with slope $m(i,j)$,
\[
d(i,j)^2 = \frac{\left [ (i-y_c) + m(i,j)(x_c-j)  \right ] ^2}{m(i,j)^2+1}
\]
Here, we use a weight function like Eq.\ref{eq1} to consider a fraction of the image
\[
D^2 = \sum_{i,j}^{N,M} w(i,j) d(i,j)^2
\]
If $(x_c, y_c)$ is the center point, $D^2$ value should be minimum.
For $x_c$
\[
\frac{d D^2}{d x_c} = \sum_{i,j}^{N,M}  \frac{m(i,j)w(i,j) [ (i-y_c) + m(i,j)(x_c-j)   ]}{m(i,j)^2+1 }= 0
\]
\[
x_c \sum_{i,j}^{N,M} \frac{w(i,j) m(i,j)^2}{m(i,j)^2+1} - y_c\sum_{i,j}^{N,M} \frac{w(i,j) m(i,j)}{m(i,j)^2+1} = \sum_{i,j}^{N,M} \frac{w(i,j) m(i,j) (m(i,j) j - i)}{m(i,j)^2+1}
\]
For $y_c$
\[
\frac{d D^2}{d y_c} = - \sum_{i,j}^{N,M} \frac{w(i,j) [ (i-y_c) + m(i,j)(x_c-j)   ]}{m(i,j)^2+1 }= 0
\]
\[
-x_c \sum_{i,j}^{N,M} \frac{w(i,j) m(i,j)}{m(i,j)^2+1} + y_c\sum_{i,j}^{N,M} \frac{w(i,j)}{m(i,j)^2+1} =- \sum_{i,j}^{N,M} \frac{w(i,j) (m(i,j) j - i)}{m(i,j)^2+1}
\]
then, we obtain the elements of the linear equation system.
\[A=\sum_{i,j}^{N,M} \frac{w(i,j) m(i,j)^2}{m(i,j)^2+1}    ,  \quad   B=-\sum_{i,j}^{N,M} \frac{w(i,j)m(i,j)}{m(i,j)^2+1}
\]
\[  
C=-\sum_{i,j}^{N,M} \frac{w(i,j) m(i,j)}{m(i,j)^2+1}  , \quad
D=\sum_{i,j}^{N,M} \frac{w(i,j)}{m(i,j)^2+1}   \]
and
\[ E=\sum_{i,j}^{N,M} \frac{w(i,j) m(i,j) (m(i,j) j - i)}{m(i,j)^2+1}   , \quad  F= - \sum_{i,j}^{N,M} \frac{w(i,j) (m(i,j) j - i)}{m(i,j)^2+1} \]
 
\[
 \left [ \begin{array}{cc}
  A & B \\
  C & D \\
 \end{array} \right ]
  \left [ \begin{array}{c}
  x_c  \\
  y_c  \\
 \end{array} \right ] = \left [ \begin{array}{c}
  E  \\
  F  \\
 \end{array} \right ]
\]
the solution is given by
\[
  \left [ \begin{array}{c}
  x_c  \\
  y_c  \\
 \end{array} \right ] = \frac{1}{AD-BC}
  \left [ \begin{array}{cc}
  D & -B \\
  -C & A \\
 \end{array} \right ] \left [ \begin{array}{c}
  E  \\
  F  \\
 \end{array} \right ]
\]
This formulation is coded in the supplementary file.
\bibliographystyle{ieeetr}
\bibliography{sample}

\newpage

\end{document}